\documentstyle[preprint,aps,floats,epsf]{revtex}

\newdimen\pmboffset
\def\oldpmb#1{\setbox0=\hbox{#1}%
\copy0\kern-\wd0 \kern\pmboffset\raise
1.732\pmboffset\copy0\kern-\wd0 \kern\pmboffset\box0}
\def\pmb#1{\mathchoice{\oldpmb{$\displaystyle#1$}}{\oldpmb{$\textstyle#1$}}
{\oldpmb{$\scriptstyle#1$}}{\oldpmb{$\scriptscriptstyle#1$}}}

\begin{document}


\preprint{CALT-68-2101}
\title{Self-Energy of Decuplet Baryons in Nuclear Matter}
\author{S.~M.~Ouellette}
\address{Lauritsen Laboratory,
California Institute of Technology,
Pasadena, CA 91125}
\author{Ryoichi~Seki}
\address{W.~K.~Kellogg Radiation Laboratory,
California Institute of Technology,
Pasadena, CA 91125\\ and\\
Department of Physics and Astronomy,
California State University,
Northridge, CA 91330}
\date{February 21, 1997}
\maketitle


\begin{abstract}
We calculate, in chiral perturbation theory, the change in the
self-energy of decuplet baryons in nuclear matter.
These self-energy shifts are relevant in studies of meson-nucleus
scattering and of neutron stars.
Our results are leading order in an expansion in powers of the ratio
of characteristic momenta to the chiral symmetry-breaking scale (or
the nucleon mass).
Included are contact diagrams generated by \mbox{4-baryon} operators,
which were neglected in earlier studies for the $\Delta$~isomultiplet
but contribute to the self-energy shifts at this order in chiral
perturbation theory.
\end{abstract}


The proper self-energy of decuplet baryons in nuclear matter is
shifted relative to the self-energy in free space by strong-%
interaction effects.
The self-energy shifts of the decuplet baryons, particularly of the
$\Delta$~isomultiplet, are relevant in studies of meson-nucleus
scattering~\cite{PionsNuclei,TheoryMeson} and of neutron stars~%
\cite{NeutronStars,Sawyer72}\@.
We calculate the self-energy shifts of the \mbox{spin-$\frac{3}{2}$}
decuplet baryons ($\Delta$, $\Sigma^*$, $\Xi^*$, $\Omega$) in nuclear
matter, using the model-independent approach of Savage and Wise~%
\cite{SavageWise96} based on chiral perturbation theory~($\chi$PT)\@.
For the $\Delta$~isomultiplet, the self-energy shifts have been
examined in various phenomenological models~\cite{Sawyer72,%
CenniDillon80,WehrbergerWittman90}\@.
These earlier calculations for the $\Delta$~isomultiplet omitted
contact diagrams necessary for a consistent momentum expansion in
$\chi$PT\@.
At leading order in $\chi$PT the relevant contact diagrams arise from
\mbox{4-baryon} operators in the chiral effective Lagrangian.
The coefficients of these operators have not yet been extracted from
experiment; however, the helicity splitting of the self-energy shifts
and the shift of the \mbox{$\Delta$-resonance} width do not depend on
these unknown coefficients.


We start from the most general effective Lagrangian consistent with the 
${\rm SU(3)}_L \times {\rm SU(3)}_R$ chiral symmetry of QCD\@.
The chiral Lagrangian is written as an expansion in derivatives and
powers of the light-quark mass matrix.
Furthermore, the decuplet and octet baryons are treated as heavy
static fields, effectively keeping only the lowest order terms in the
chiral Lagrangian expanded in powers of the baryon momenta over baryon
masses.
These expansions are justified when the characteristic momentum scale
of the processes considered is small compared to the chiral
symmetry-breaking scale and the baryon masses.


The lowest order in a density expansion for nuclear matter is a Fermi
gas of non-interacting protons and neutrons with Fermi momenta
$p_{F,p}^{}$ and $p_{F,n}^{}$\@.
In this framework, the characteristic momentum scales relevant in the
derivative expansion are $p_{F,p}^{}$, $p_{F,n}^{}$, and the decuplet
baryon 3-momentum, $k = |\vec{k}|$\@.
Since the density of a degenerate Fermi gas is given by
$d=p_F^3/3\pi^2$, the density expansion for nuclear matter is
consistent with the chiral derivative expansion.
The presence of the nuclear medium modifies the nucleon propagator
through two effects, the presence of nucleon-hole intermediate states
and the nucleon intermediate states inaccessible due to
Pauli-blocking.
The heavy-nucleon propagator in nuclear matter with Fermi momentum
$p_F^{}$ is then
\begin{equation}
\label{eq:NucProp}
	\frac{i\theta(p_F^{} - |\vec{p}|)}{p^0 - i\epsilon}
	+\frac{i\theta(|\vec{p}| - p_F^{})}{p^0 + i\epsilon}
\end{equation}
at lowest order in the nuclear density.


The decuplet baryon self-energy in nuclear matter $E_{NM}(\vec{k})$ is
the location of the pole in the \mbox{2-point} function, given as the
solution of
\begin{equation}
\label{eq:Pole}
	E_{NM}(\vec{k}) - (m_T-m_B) 
		- \Sigma_{NM}(E_{NM}(\vec{k}),\vec{k}) = 0\ ,
\end{equation}
where $(m_T-m_B)$ is the mass appearing in the heavy-field Lagrangian,
Eq.~(\ref{eq:Lchi}) below, and $\Sigma_{NM}$ is the sum of connected
one-particle irreducible diagrams in the expansion of the \mbox{2-point}
function.
We evaluate the self-energy shift,
\( \delta\Sigma(k) \equiv E_{NM}(\vec{k}) - E_{VAC}(\vec{k}) \),
by expanding the nuclear-matter self-energy about the pole of the
free-space \mbox{2-point} function
\begin{eqnarray}
\label{eq:DefShift}
	\delta\Sigma(k) & = &
		\left[\Sigma_{NM}(E_{VAC}(\vec{k}),\vec{k})
		- \Sigma_{VAC}(E_{VAC}(\vec{k}),\vec{k})\right]
		\nonumber\\*
	& & {} + \left( E_{NM}(\vec{k})-E_{VAC}(\vec{k})\right)
		\left.\frac{\partial}{\partial p^0}
		\Sigma_{NM}(p^0,\vec{k})\right|_{p^0=E_{VAC}(\vec{k})}
		+ \cdots\ .
\end{eqnarray}
Note that only diagrams with an internal nucleon propagator contribute
to the first term of this expansion, and that following terms are
higher order in $\chi$PT\@.
The real part of the self-energy shift,
$\delta\Pi(k) \equiv \mbox{Re}[\delta\Sigma(k)]$,
modifies the baryon energy-momentum dispersion relation from the
free-space dispersion relation.
The change in the resonance width is given by
$\delta\Gamma(k) = -2\,\mbox{Im}[\delta\Sigma(k)]$\@.


Because the nuclear medium we consider is rotationally invariant, the
only preferred direction is along the decuplet baryon \mbox{3-momentum},
$\vec{k}$\@.
Symmetry under rotations about $\vec{k}$ ensures that the self-energy
shifts do not mix baryon states with different spin projections along
the direction of $\vec{k}$; i.e., the self-energy shifts of the
\mbox{spin-$\frac{3}{2}$} decuplet baryons are diagonal in the baryon
helicity states.
Parity invariance of the strong interaction, which is manifest in the
chiral effective Lagrangian, dictates that the self-energy shifts do
not depend on the sign of the helicity.
In the limit of vanishing decuplet \mbox{3-momentum}, the rotational
symmetry is elevated to full SU(2) rotational invariance, and the
self-energy shifts must be independent of the decuplet spin projection
along any direction.
In this limit we expect the helicity splitting of the self-energy
shifts to vanish;
\( \delta\Sigma(0)|{{}_{h = \pm\frac{3}{2}}^{}}
	- \delta\Sigma(0)|{{}_{h = \pm\frac{1}{2}}^{}} = 0 \)\@.
Consequently, at each order of $\chi$PT, if the self-energy shifts
are momentum independent, the shifts are also helicity independent.


In the effective Lagrangian of $\chi$PT, the pseudo-Goldstone bosons
($\pi$, $K$, $\eta$) are organized as a $3 \times 3$ special unitary
matrix $\pmb{\Sigma} = \exp(2i\pmb{\pi}/f)$, where $f$ is equal to the
pion decay constant ($f_\pi \simeq 132$ Mev) at lowest order in
$\chi$PT and
\begin{equation}
\label{eq:PiField}
	\pmb{\pi} = \left[\begin{array}{ccc}
		\pi^0/\sqrt{2} + \eta/\sqrt{6} & \pi^+ & K^+ \\
		\pi^- & -\pi^0/\sqrt{2} + \eta/\sqrt{6} & K^0 \\
		K^- & \overline{K^0} & -\sqrt{\frac{2}{3}}\eta
		\end{array}\right]\ .
\end{equation}
The interactions of the pseudo-Goldstone bosons with baryon fields are
formulated in terms of the vector and axial-vector chiral fields
\begin{eqnarray}
\label{eq:VAField}
	V_\mu & = & \frac{1}{2}(\xi^\dag\partial_\mu\xi
		+ \xi\partial_\mu\xi^\dag) \ ,\nonumber\\*
	A_\mu & = & \frac{i}{2}(\xi^\dag\partial_\mu\xi
		- \xi\partial_\mu\xi^\dag) \ ,
\end{eqnarray}
where $\xi \equiv \sqrt{\pmb{\Sigma}} = \exp(i\pmb{\pi}/f)$\@.


To lowest order in the derivative expansion, the chiral effective
Lagrangian quadratic in decuplet baryon operators is
\begin{eqnarray}
\label{eq:InitialL}
	{\cal L}_0 & = & 
		-\overline{T_0}{{}_\mu^{}}{{}_{}^{abc}}
		(i\gamma{{}_{}^\nu} {\cal D}{{}_\nu^{}} - m_T) 
		T_0{{}_{}^\mu}{{}_{abc}^{}}
		+ {\cal H}\overline{T_0}{{}_\mu^{}}{{}_{}^{abc}}
		\gamma{{}_\nu^{}}\gamma{{}_5^{}}
		T_0{{}_{}^\mu}{{}_{abd}^{}}
		A{{}_{}^\nu}{{}_c^d} \nonumber\\*
	& & {} + {\cal C}(\epsilon{{}_{abc}^{}}
		\overline{T_0}{{}_\mu^{}}{{}_{}^{ade}}
		A{{}_{}^\mu}{{}_d^b} B_0{{}_e^c}
		+ \epsilon{{}_{}^{abc}} \overline{B_0}{{}_c^e}
		A{{}_\mu^{}}{{}_b^d} T_0{{}_{}^\mu}{{}_{ade}^{}})\ ,
\end{eqnarray}
where ($\mu$,$\nu$) are Lorentz indices, ($a$--$e$) are chiral SU(3)
indices, and Dirac spinor indices are suppressed.
$T_0{{}_{}^\mu}{{}_{abc}^{}}$ is a Rarita-Schwinger field which
transforms under chiral SU(3) as a completely-symmetric \mbox{rank-3}
tensor and represents the decuplet fields as follows:
\begin{equation}
\label{eq:TField}
	\begin{array}{lll}
	T_0{{}_{}^\mu}{{}_{111}^{}} = \Delta^{++},&
	T_0{{}_{}^\mu}{{}_{113}^{}} = \Sigma^{*+}/\sqrt{3},&
	T_0{{}_{}^\mu}{{}_{133}^{}} = \Xi^{*0}/\sqrt{3},\\
	T_0{{}_{}^\mu}{{}_{112}^{}} = \Delta^+/\sqrt{3},\mbox{\ \ \ }&
	T_0{{}_{}^\mu}{{}_{123}^{}} = \Sigma^{*0}/\sqrt{6},\mbox{\ \ \ }&
	T_0{{}_{}^\mu}{{}_{233}^{}} = \Xi^{*-}/\sqrt{3},\\
	T_0{{}_{}^\mu}{{}_{122}^{}} = \Delta^0/\sqrt{3},&
	T_0{{}_{}^\mu}{{}_{223}^{}} = \Sigma^{*-}/\sqrt{3},&
	T_0{{}_{}^\mu}{{}_{333}^{}} = \Omega^-,\\*
	T_0{{}_{}^\mu}{{}_{222}^{}} = \Delta^-.& &
	\end{array}
\end{equation}
$B_0{{}_a^b}$ represents the hyperon octet as a $3 \times 3$ matrix of
Dirac fields
\begin{equation}
\label{eq:BField}
	B_0 = \left[\begin{array}{ccc}
	\Sigma^0/\sqrt{2} + \Lambda/\sqrt{6} & \Sigma^+ & p \\
	\Sigma^- & -\Sigma^0/\sqrt{2} + \Lambda/\sqrt{6} & n \\
	\Xi^- & \Xi^0 & -\sqrt{\frac{2}{3}}\Lambda
	\end{array}\right]\ .
\end{equation}
The action of the chiral covariant derivative ${\cal D}{{}_\mu^{}}$ on
the baryon fields is given by
\begin{eqnarray} 
\label{eq:DonB}
	({\cal D}{{}_\mu^{}} B_0){{}_a^b} & = &
		\partial{{}_\mu^{}} B_0{{}_a^b}
		+ V{{}_\mu^{}}{{}_a^d} B_0{{}_d^b}
		- V{{}_\mu^{}}{{}_d^b} B_0{{}_a^d} \nonumber\\*
	& = & (\partial{{}_\mu^{}} B_0 + [V{{}_\mu^{}}, B_0]){{}_a^b}
		\ ,\nonumber\\
\label{eq:DonT}
	({\cal D}{{}_\mu^{}} T_0{{}_{}^\nu}){{}_{abc}^{}} & = &
		\partial{{}_\mu^{}} T_0{{}_{}^\nu}{{}_{abc}^{}}
		+ V{{}_\mu^{}}{{}_a^d} T_0{{}_{}^\nu}{{}_{dbc}^{}}
		+ V{{}_\mu^{}}{{}_b^d} T_0{{}_{}^\nu}{{}_{adc}^{}}
		+ V{{}_\mu^{}}{{}_c^d} T_0{{}_{}^\nu}{{}_{abd}^{}}\ .
\end{eqnarray}


Following Jenkins and Manohar~\cite{JenkinsManohar91}, we obtain a
consistent derivative expansion by treating the baryons in the
heavy-fermion formalism.
We rewrite the chiral effective Lagrangian treating both octet and
decuplet baryons as heavy static fields $T{{}_{}^\mu}$ and $B$
defined by
\begin{eqnarray}
\label{eq:HeavyTB}
	T{{}_{}^\mu}(x) & = & \exp(im_Bt) T_0{{}_{}^\mu}(x)
		\ ,\nonumber\\*
	B(x) & = & \exp(im_Bt) B_0(x) \ ,
\end{eqnarray}
where $m_B$ is the octet baryon mass, and with the additional
constraints $\gamma_0 T{{}_{}^\mu} = +T{{}_{}^\mu}$ and
$\gamma_0 B = +B$\@.
Replacing the Dirac structure of the Lagrangian with \mbox{2-component}
Pauli spinors, and keeping only terms at lowest order in $1/m_B$ and
$1/m_T$, the chiral effective Lagrangian is
\begin{eqnarray}
\label{eq:Lchi}
	{\cal L} & = & T^\dag{{}_{}^{(j)}}{{}_{}^{abc}}
		(i{\cal D}{{}_0^{}} - (m_T - m_B))
		T{{}_{}^{(j)}}{{}_{abc}^{}}
		+ {\cal H} T^\dag{{}_{}^{(j)}}{{}_{}^{abc}}
		\sigma{{}_{}^{(k)}}
		T {{}_{}^{(j)}}{{}_{abd}^{}} A{{}_{}^{(k)}}{{}_c^d}
		\nonumber\\*
	& & {} + {\cal C}(\epsilon{{}_{abc}^{}}
		T^\dag{{}_{}^{(j)}}{{}_{}^{ade}}
		A{{}_{}^{(j)}}{{}_d^b} B{{}_e^c} 
		+ \epsilon{{}_{}^{abc}}
		B^\dag{{}_c^e} A{{}_{}^{(j)}}{{}_b^d}
		T{{}_{}^{(j)}}{{}_{ade}^{}})\ .
\end{eqnarray}
In the $1/m_B$, $1/m_T$ expansion, the Lorentz \mbox{4-vectors} have
been replaced by spacial vectors with indices ($j$,$k$) implicitly
summed over 1--3, and the spinor indices have been suppressed.


We adapt the power-counting rules of Savage and Wise~%
\cite{SavageWise96} to accommodate the decuplet baryon mass $m_T$
and 3-momentum $k$.  Both octet and decuplet masses, $m_T$ and $m_B$,
are treated as the same order as the chiral symmetry breaking scale
$\Lambda_\chi \sim 4\pi f$\@.
The pseudo-Goldstone boson masses ($m_\pi$, $m_K$, $m_\eta$) provide a
smaller mass scale, denoted by $q$\@.
The decuplet-octet mass splitting $m_T-m_B$ and the characteristic
momenta $p_{F,p}^{}$, $p_{F,n}^{}$, and $k$ are also considered to be
of order $q$\@.
(The light quark masses in the symmetry-breaking Lagrangian are order
$q^2$\@.)
With the exception of Weinberg's notorious infrared-divergent
diagrams~\cite{Weinberg90}, a Feynman diagram with $L$ loops and $V_i$
vertices with $n_i$ baryon operators and $d_i$ derivatives (or factors
of pseudo-Goldstone boson masses), contributes to the self-energy of
the baryon field at order $q(q/\Lambda_\chi)^\alpha$ where
\begin{equation}
\label{eq:ChiralOrder}
	\alpha = 2L+\sum_i V_i(d_i+\frac{1}{2}n_i-2)\ .
\end{equation}
There are two kinds of diagrams that contribute in the leading order
where $\alpha = 2$:
1)~meson-nucleon loop diagrams where $n_i = 2$ and $d_i = 1$ (shown in
Fig.~1a), and
2)~contact diagrams (shown in Fig.~1b), with one insertion of an
operator with $n_i=4$ and $d_i=0$ containing two decuplet baryon
fields and two octet baryon fields.


\begin{figure}[b]
\centerline{\unitlength=1cm \makebox(0,2)[t]{(a)}
	\epsfysize=2cm \epsfbox{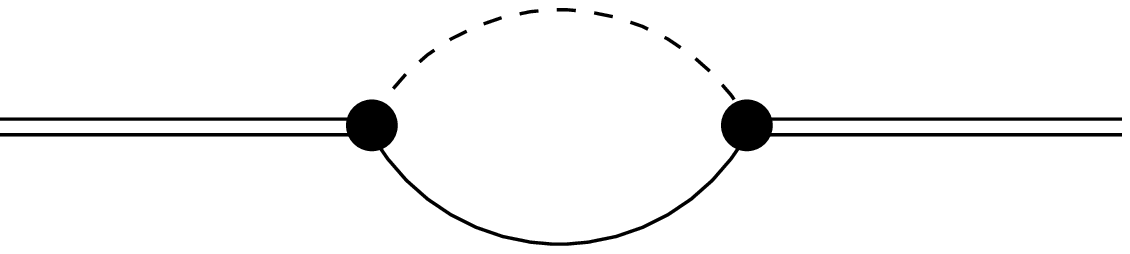} \hspace{\fill}
	\makebox(0,2)[t]{(b)} \epsfysize=2cm \epsfbox{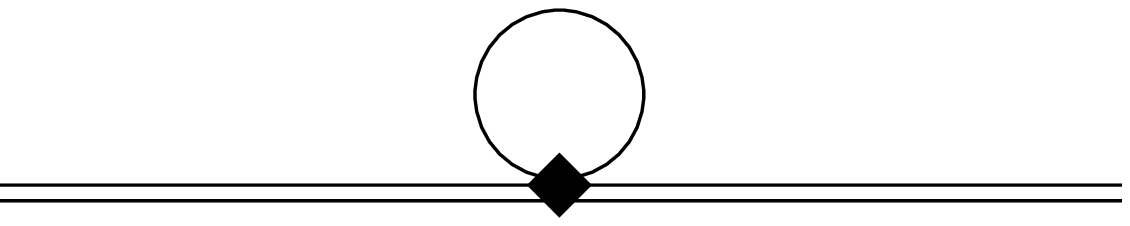}}
\vspace{\baselineskip}
\caption{Diagrams contributing to self-energy shifts at lowest
order in $\chi$PT:  (a)~meson-nucleon loop diagram and (b)~contact
diagram.  Double solid lines represent decuplet baryon fields, single
solid lines represent nucleon fields, and dashed lines represent
pseudo-Goldstone boson fields.}
\end{figure}


The meson-nucleon loop diagrams are generated by terms in
Eq.~(\ref{eq:Lchi}).  For the contact diagrams, we construct the
relevant \mbox{4-baryon} operators by considering the spin and chiral
structures of $T{{}_{}^{(j)}}B$\@.
The operator product $T{{}_{}^{(j)}}B$ decomposes under rotational
SU(2) as \( {\bf \frac{3}{2} \otimes \frac{1}{2}} = {\bf 2 \oplus 1 } \)
and decomposes under chiral SU(3) as
\( {\bf 10 \otimes 8 } = {\bf 35 \oplus 27 \oplus 10 \oplus 8 } \)\@.
Therefore the chiral effective Lagrangian contains eight linearly
independent \mbox{4-baryon} operators of the form
$T^\dag{{}_{}^{(j)}}T{{}_{}^{(k)}}B^\dag B$, which contribute to the
self-energy shift at leading order.
We choose to write the Lagrangian containing these \mbox{4-baryon}
operators as
\begin{eqnarray}
\label{eq:dLchi}
	\delta{\cal L} & = & {}
		- \frac{d_1}{f^2}[T^\dag{{}_{}^{(j)}}{{}_{}^{abc}}
		T{{}_{}^{(j)}}{{}_{abc}^{}}][B^\dag{{}_d^e}B{{}_e^d}]
		- \frac{d_5}{f^2}[T^\dag{{}_{}^{(j)}}{{}_{}^{abc}}
		\sigma{{}_{}^{(k)}}T{{}_{}^{(j)}}{{}_{abc}^{}}]
		[B^\dag{{}_d^e}\sigma{{}_{}^{(k)}}B{{}_e^d}]
		\nonumber\\*
	& & {} - \frac{d_2}{f^2}[T^\dag{{}_{}^{(j)}}{{}_{}^{abc}}
		T{{}_{}^{(j)}}{{}_{abd}^{}}][B^\dag{{}_e^d}B{{}_c^e}]
		- \frac{d_6}{f^2}[T^\dag{{}_{}^{(j)}}{{}_{}^{abc}}
		\sigma{{}_{}^{(k)}}T{{}_{}^{(j)}}{{}_{abd}^{}}]
		[B^\dag{{}_e^d}\sigma{{}_{}^{(k)}}B{{}_c^e}]
		\nonumber\\*
	& & {} - \frac{d_3}{f^2}[T^\dag{{}_{}^{(j)}}{{}_{}^{abc}}
		T{{}_{}^{(j)}}{{}_{abd}^{}}][B^\dag{{}_c^e}B{{}_e^d}]
		- \frac{d_7}{f^2}[T^\dag{{}_{}^{(j)}}{{}_{}^{abc}}
		\sigma{{}_{}^{(k)}}T{{}_{}^{(j)}}{{}_{abd}^{}}]
		[B^\dag{{}_c^e}\sigma{{}_{}^{(k)}}B{{}_e^d}]
		\nonumber\\*
	& & {} - \frac{d_4}{f^2}[T^\dag{{}_{}^{(j)}}{{}_{}^{abc}}
		T{{}_{}^{(j)}}{{}_{ade}^{}}][B^\dag{{}_b^d}B{{}_c^e}]
		- \frac{d_8}{f^2}[T^\dag{{}_{}^{(j)}}{{}_{}^{abc}}
		\sigma{{}_{}^{(k)}}T{{}_{}^{(j)}}{{}_{ade}^{}}]
		[B^\dag{{}_b^d}\sigma{{}_{}^{(k)}}B{{}_c^e}]
		\ ,
\end{eqnarray}
where square brackets indicate sums over spinor indices.
Because the baryon helicity is conserved in the self-energy diagrams,
the terms in Eq.~(\ref{eq:dLchi}) of the form 
\( [T^\dag{{}_{}^{(j)}}\sigma{{}_{}^{(k)}}T{{}_{}^{(j)}}]
	[B^\dag\sigma{{}_{}^{(k)}}B] \)
do not generate contact diagrams contributing to the self-energy shifts.


To account for the leading effects of the symmetry-breaking terms in
the effective Lagrangian, we use non-zero masses $m_\pi$, $m_K$, and
$m_\eta$ for the pseudo-Goldstone bosons and we use isospin multiplet
masses in place of chiral multiplet masses for the baryons.
For the decuplet baryons, the effective masses replacing $m_T$ are
$m_\Delta$, $m_{\Sigma^*}$, $m_{\Xi^*}$, and $m_\Omega$\@.
We use the nucleon mass $m_N$ for the octet baryon mass $m_B$ since
diagrams containing the other octet baryons cancel between nuclear
matter and free space.


From the chiral effective Lagrangians given in Eqs.~(\ref{eq:Lchi})
and~(\ref{eq:dLchi}), we now derive the self-energy shifts of the
decuplet baryons ($\Delta$, $\Sigma^*$, $\Xi^*$, $\Omega$) in nuclear
matter to leading order in chiral perturbation theory.
Here, we have introduced the convenient mass scales,
\begin{eqnarray} 
\label{eq:DeltaScale}
	\mu = & \sqrt{(m_\Delta - m_N )^2 - m_\pi{}^2} 
		& \simeq \mbox{ 255 MeV,} \nonumber \\*
\label{eq:SigmaScale}
	\tilde{\mu} = & \sqrt{{m_K}^2 - (m_{\Sigma^*} - m_N)^2}
		\mbox{\ } & \simeq \mbox{ 215 MeV.}
\end{eqnarray}
For the member of each isomultiplet with lowest \mbox{3-component} of
isospin ($\Delta^-$, $\Sigma^{*-}$, $\Xi^{*-}$, $\Omega^-$), we obtain
\begin{eqnarray}
\label{eq:DeltaShift}
	\delta\Pi_{\Delta^-} & = &
		\frac{p_{F,n}^3}{9\pi^2f^2}(3d_1+3d_2)
		+\frac{p_{F,p}^3}{9\pi^2f^2}(3d_1)
		+\frac{|{\cal C}|^2}{(4\pi f)^2}\left\{
		\frac{2\mu^3}{3}\ln\left|\frac{(p_{F,n}^{}-\mu)^2-k^2}
		{(p_{F,n}^{}+\mu)^2-k^2}\right|\right.
		\nonumber\\*
	& & \mbox{}+\frac{4p_{F,n}^{}}{9}(2p_{F,n}^2+3\mu^2)
		-\frac{\mu^2}{3k}(p_{F,n}^2-\mu^2-k^2)
		\ln\left|\frac{(p_{F,n}-k)^2-\mu^2}
		{(p_{F,n}^{}+k)^2-\mu^2}\right|
		\nonumber\\*
	& & \mbox{}\pm\left[\frac{p_{F,n}^{}}{6k^2}
		(p_{F,n}^2-\mu^2-k^2)^2-\frac{p_{F,n}^{}}{9}
		(p_{F,n}^2+3\mu^2+3k^2)+\frac{1}{24k^3}
		(p_{F,n}^2-\mu^2-k^2)\right.
		\nonumber\\*
	& & \left.\left.\mbox{}\times
		((p_{F,n}^{}-k)^2-\mu^2)((p_{F,n}^{}+k)^2-\mu^2)
		\ln\left|\frac{(p_{F,n}^{}-k)^2-\mu^2}
		{(p_{F,n}^{}+k)^2-\mu^2}\right|\ \right]\right\}
		\ ,\\
\label{eq:SigmaShift}
	\delta\Pi_{\Sigma^{*-}} & = &
		\frac{p_{F,n}^3}{9\pi^2f^2}(3d_1+2d_2+d_3+d_4)
		+\frac{p_{F,p}^3}{9\pi^2f^2}(3d_1+d_3)
		\nonumber\\*
	& & \mbox{}+\frac{|{\cal C}|^2}{(4\pi f)^2}\left\{
		\frac{4\tilde{\mu}^3}{9}\arctan\left(
		\frac{p_{F,n}^{}-k}{\tilde{\mu}}\right)
		+\frac{4\tilde{\mu}^3}{9}\arctan\left(
		\frac{p_{F,n}^{}+k}{\tilde{\mu}}\right)\right.
		\nonumber\\*
	& & \mbox{}+\frac{4p_{F,n}^{}}{27}(2p_{F,n}^2-3\tilde{\mu}^2)
		+\frac{\tilde{\mu}^2}{9k}(p_{F,n}^2+\tilde{\mu}^2-k^2)
		\ln\left(\frac{(p_{F,n}^{}-k)^2+\tilde{\mu}^2}
		{(p_{F,n}^{}+k)^2+\tilde{\mu}^2}\right)
		\nonumber\\*
	& & \mbox{}\pm\left[\frac{p_{F,n}^{}}{18k^2}
		(p_{F,n}^2+\tilde{\mu}^2-k^2)^2-\frac{p_{F,n}^{}}{27}
		(p_{F,n}^2-3\tilde{\mu}^2+3k^2)+\frac{1}{72k^3}
		(p_{F,n}^2+\tilde{\mu}^2-k^2)\right.
		\nonumber\\*
	& & \left.\left.\mbox{}\times
		((p_{F,n}^{}-k)^2+\tilde{\mu}^2)
		((p_{F,n}^{}+k)^2+\tilde{\mu}^2)
		\ln\left(\frac{(p_{F,n}^{}-k)^2+\tilde{\mu}^2}
		{(p_{F,n}^{}+k)^2+\tilde{\mu}^2}\right)
		\ \right]\right\}
		\ ,\\
\label{eq:XiShift}
	\delta\Pi_{\Xi^{*-}} & = & 
		\frac{p_{F,n}^3}{9\pi^2f^2}(3d_1+d_2+2d_3+d_4)
		+\frac{p_{F,p}^3}{9\pi^2f^2}(3d_1+2d_3)\ ,\\
\label{eq:OmegaShift}
	\delta\Pi_{\Omega^-} & = & 
		\frac{p_{F,n}^3}{9\pi^2f^2}(3d_1+3d_3)
		+\frac{p_{F,p}^3}{9\pi^2f^2}(3d_1+3d_3)\ .
\end{eqnarray}
In Eqs.~(\ref{eq:DeltaShift}) and~(\ref{eq:SigmaShift}), and in
Eq.~(\ref{eq:DeltaWidth}) below, the plus sign of `$\pm$' corresponds
to the $h=\pm\frac{3}{2}$ helicity states and the minus sign
corresponds to the $h=\pm\frac{1}{2}$ helicity states.
There is no helicity splitting of the self-energy for $\Xi^{*+}$,
$\Xi^{*-}$, or $\Omega^-$~baryons because the chiral Lagrangian does
not couple these baryons to a nucleon and single pseudo-Goldstone
boson.
At this order in $\chi$PT the resonance width in nuclear matter is
shifted only for the $\Delta$~isomultiplet.
For the shift of the width of the $\Delta^-$, we find
\begin{eqnarray}
\label{eq:DeltaWidth}
	\delta\Gamma_{\Delta^-} & = &
		\frac{|{\cal C}|^2}{(4\pi f)^2}\left\{
		-\frac{8\pi\mu^3}{3}\Theta(p_{F,n}^{}-\mu-k)
		+\Theta(p_{F,n}^{}-\mu+k)\Theta(\mu^2-(p_{F,n}^{}-k)^2)
		\right.\nonumber\\*
	& & \left.\mbox{}\times\frac{2\pi}{3k}(p_{F,n}^2-(\mu-k)^2)
		\left[-\mu^2\pm\frac{1}{8k^2}(p_{F,n}^2-(\mu+k)^2)
		(p_{F,n}^2-\mu^2-k^2)\right]\right\}\ .
\end{eqnarray}


For the member of each isomultiplet with the highest
\mbox{3-component} of isospin ($\Delta^{++}$, $\Sigma^{*+}$,
$\Xi^{*0}$, $\Omega^-$), the self-energy shifts are obtained from the
results for the corresponding isomultiplet member with lowest
\mbox{3-component} of isospin by exchanging $p_{F,p}^{}$ and
$p_{F,n}^{}$ in \mbox{Eqs. (\ref{eq:DeltaShift})--%
(\ref{eq:DeltaWidth})}\@.
The self-energy shifts of the remaining decuplet baryons are given by
the following relations:
\begin{eqnarray}
\label{eq:DeltaPlus}
	\delta\Sigma_{\Delta^+} & = & 
		\frac{2}{3}\delta\Sigma_{\Delta^{++}}
		+\frac{1}{3}\delta\Sigma_{\Delta^-}\ ,\\
\label{eq:DeltaZero}
	\delta\Sigma_{\Delta^0} & = & 
		\frac{1}{3}\delta\Sigma_{\Delta^{++}}
		+\frac{2}{3}\delta\Sigma_{\Delta^-}\ ,\\
\label{eq:SigmaZero}
	\delta\Sigma_{\Sigma^{*0}} & = &
		\frac{1}{2}\delta\Sigma_{\Sigma^{*+}}
		+\frac{1}{2}\delta\Sigma_{\Sigma^{*-}}\ .
\end{eqnarray}


The leading corrections to our results are due to infrared-divergent
\mbox{2-loop} diagrams.
The infrared-divergent diagrams are regulated by insertions of the
baryon kinetic energy, resulting in powers of the baryon mass in the
numerator of the diagrams.
The contributions of the infrared-divergent \mbox{2-loop} diagrams are
suppressed by a factor of order $qm_B/\Lambda_\chi{}^2$\@.
Whether these corrections can be considered small depends strongly on
numerical factors arising from the loop integrations.
Diagrams with more than one loop which are not infrared-divergent, and
diagrams with insertions of the baryon kinetic energy, are suppressed
by $(q/\Lambda_\chi)^2$ relative to our results.
Expanding the self-energy in nuclear matter beyond lowest order about
the pole of the free-space \mbox{2-point} function, also gives
corrections suppressed by $(q/\Lambda_\chi)^2$\@.
Diagrams generated by terms in the general chiral effective Lagrangian
with more than four baryon operators are suppressed by
$(q/\Lambda_\chi)^3$\@.


Our work differs in two ways from earlier calculations of the
\mbox{$\Delta$-isomultiplet} self-energies in nuclear matter~%
\cite{Sawyer72,CenniDillon80,WehrbergerWittman90}\@.
We use chiral SU(3) symmetry to extend the calculation to the entire
\mbox{spin-$\frac{3}{2}$} baryon decuplet; and for the \mbox{$\Delta$-%
isomultiplet} self-energy shifts, we find new momentum-independent
contributions from contact diagrams generated by \mbox{4-baryon}
operators.
A comprehensive discussion of the momentum dependence of the results
for the $\Delta$~isomultiplet, without the contact diagram
contributions, is given by Cenni and Dillon~\cite{CenniDillon80}\@.
Of particular interest is their discussion of the logarithmic
divergence of $\delta\Pi_\Delta(k)$ as $k\rightarrow0$ when
$p_F^{}=\mu\simeq\mbox{255 MeV}$ in the context of coupled $\Delta$~%
and $N\pi$~eigenmodes.
The \mbox{$\Sigma^*$-isomultiplet} self-energy shifts do not have a
similar divergence because the $\Sigma^* \rightarrow NK$ decay is
kinematically forbidden for any $p_{F,p}^{}$ and $p_{F,n}^{}$\@.
The contact diagrams included here offset the decuplet baryon
self-energy shifts by terms proportional to $p^3_{F,p}$ and
$p^3_{F,n}$\@.


The dimensionless coefficients $d_i$ of the contact terms in the
chiral effective Lagrangian, Eq.~(\ref{eq:dLchi}), have not yet been
extracted from experiment.
Because the coefficients $d_i$ are unknown, we do not evaluate the
self-energy shifts of the decuplet baryons numerically.
Also, it is unclear whether the self-energy shift enhances or
suppresses the decuplet baryon populations in dense nuclear matter.
Note however, the helicity splittings of the self-energy shifts of the
$\Delta$~and $\Sigma^*$~isomultiplets and the shift of the 
\mbox{$\Delta$-resonance} width do not depend on the coefficients
$d_i$\@.
Furthermore, the \mbox{$\Delta$-isomultiplet} self-energy shifts
depend only on the coefficients $d_1$ and $d_2$ due to constraints
from chiral SU(3) symmetry.


Finally, we note that although the coefficients $d_i$ of the
\mbox{4-baryon} operators in the chiral effective Lagrangian are not
yet known, we can reduce the number of independent parameters from
eight to two by appealing to SU(6) spin-flavor symmetry~%
\cite{KaplanSavage96}\@.
Under spin-flavor SU(6), the coefficients $d_i$ are determined by
\begin{equation}
\label{eq:KaplanSavage}
	\begin{array}{ll}
	d_1 = 2a+\frac{5}{9}b,\mbox{\ \ \ } & d_2 = -\frac{5}{9}b, \\*
	d_3 = -\frac{5}{9}b, & d_4 = -\frac{2}{9}b, \\
	d_5 = \frac{1}{9}b, & d_6 = -\frac{1}{3}b, \\*
	d_7 = -\frac{1}{9}b, & d_8 = -\frac{2}{9}b,
	\end{array}
\end{equation}
in terms of the coefficients $a$ and $b$ defined by Kaplan and Savage.


In conclusion, we have determined the self-energy shifts of the
\mbox{spin-$\frac{3}{2}$} decuplet baryons in nuclear matter to
leading order in chiral perturbation theory.
We find there are momentum-independent contributions which depend on
four of the coefficients $d_i$ in the chiral effective Lagrangian
which are not yet known.
However, the change in the \mbox{$\Delta$-resonance} width and the
helicity splittings of the self-energy shifts are independent of these
unknown coefficients.
The validity of the leading order results depends critically on the
size of corrections due to regulated infrared-divergent Feynman
diagrams at \mbox{2-loop} order.
We hope that in future work the relative importance of the leading
corrections to our results will be determined.


We would like to thank Mark~Wise for guidance and instructive
discussions throughout the course of this work.
S.~M.~O. would also like to thank A.K.~Leibovich and I.~Stewart for
encouragement and useful comments on the manuscript.
This work is supported in part by the U.S.~DOE
(\mbox{DE-FG03-92-ER40701}) and the U.S.~NSF (\mbox{PHY94-12818} and
\mbox{PHY94-20470}) at Caltech, and by the U.S.~DOE
(\mbox{DE-FG03-87ER40347}) at CSUN\@.


\end{document}